\documentclass[a4]{article}
\usepackage[a4paper, total={6.5in, 9in}]{geometry}

\usepackage{moreverb}
\usepackage{multirow}
\usepackage{graphicx}
\usepackage[colorlinks,bookmarks=false,citecolor=red,urlcolor=red]{hyperref}
\usepackage{amsfonts,amssymb,amsbsy,amsmath,amsthm}

\begin{document}


\title{Boussinesq-like problems in discrete media}

\maketitle

\begin{center}
	\author{I.~G.~Tejada\footnote{Dpto. Ing. y M. del Terreno, Universidad Polit\'ecnica de Madrid. ignacio.gtejada (at) upm.es}}
\end{center}

\begin{abstract}
		Vertical loads acting on the surface of a half-space made of discrete and elastic particles are supported by a network of force chains that changes with the specific realization of the packing. These force chains can be transformed into equivalent stress fields, but the obtained values are usually different to those expected from the solution of the corresponding boundary value problem. In this research the relationship between discrete and continuum approaches to Boussinesq-like problems is explored in the light of classical statistical mechanics. In principal directions, the anticipated statistical distributions of the extensive stress (\textit{i.e.} the product of the stress by the volume) are exponential distributions for normal components and Laplace distributions for shear components.  The parameters scaling these distributions can be obtained from the solutions provided by continuum approaches in most of the cases. This has been validated through massive numerical simulation with the discrete element method. These results could be of interest in highly fragmented, faulted or heterogeneous media or for small length scales.
\end{abstract}
	
\section{Introduction}
\label{intro}
\begin{figure}
		\centering
		\includegraphics[width=0.45\textwidth]{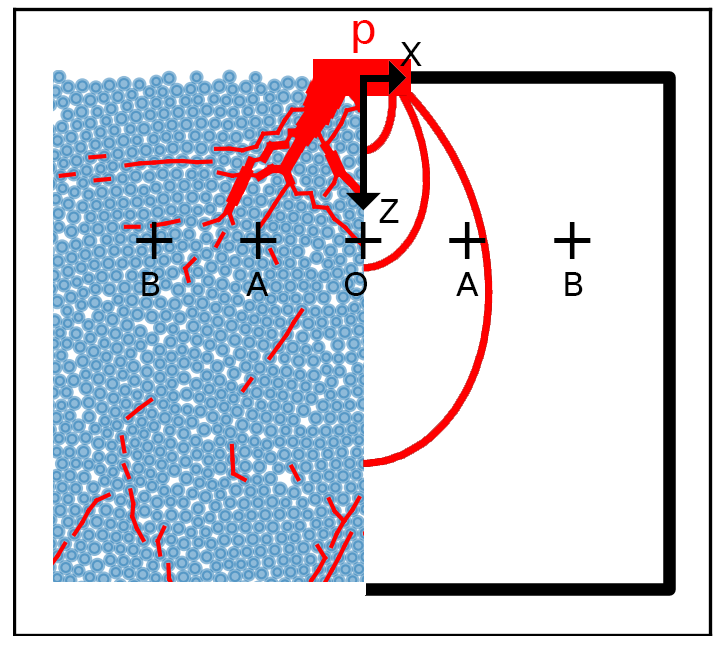}
		\caption{The solution of a Boussinesq-like problem in discrete and continuum media. In the former, the equilibrium is described by a network of force chains. In the latter by an stress field whose iso-stress points form stress bulbs. O, A and B are the control points used in this research.}
		\label{Fig:DiscreteVsCont}       
	\end{figure}

	The estimation of the stresses caused in the ground by surface loads is one of the most known problems in geotechnics. If the ground is supposed to be a continuous, homogenous isotropic and linear elastic half-space, the solution for the case of a vertical point force was given by Boussinesq~\cite{Boussinesq1885}. The two dimensional version (\textit{i.e.} a vertical line load acting on the surface) was solved a few years later by Flamant~\cite{Flamant1892}. When point forces are replaced by surface loads, solutions can be obtained from the superposition of infinitesimal loading states (more examples in~\cite{verruijt2007}). The situations in which there is a vertical load acting on the surface of a half-space are referred to as Boussinesq-like problems in this work.\\
	From the point of view of continuum mechanics, the stress field in Boussinesq-like conditions is obtained by solving the equations governing the corresponding linear elastic boundary value problem. These equations include three tensor partial differential equations for the balance of linear momentum and six infinitesimal strain-displacement relations. The system of differential equations is completed by a set of linear algebraic constitutive relations (Hooke's law). For example, in the case of 2D and a finite surface load, the line that connects all points below the ground surface objected to the same vertical pressure is a well-known stress bulb (Fig.~\ref{Fig:DiscreteVsCont}).\\ 
	When the half-space is not homogeneous, a more detailed geometry can be used in the model. If the material is not isotropic or it is not linear elastic, and advanced constitutive relationship can be used~\cite{Lade05}. However if the half-space is not a continuous body but a dense packing of distinct and elastic particles, there is no clear theoretical framework to solve the geotechnical problem. Experience has shown that in such circumstances the surface load is supported by a system of interparticle forces, which organize in force chains and result in an  inhomogeneous transmission of stresses (some particles are highly loaded while others are not, Fig.~\ref{Fig:DiscreteVsCont}). This  was initially observed in photoelastic experiments  with particles made of a birefringent material \cite{Drescher72}. The volumetric average of the stress field within any particle can be determined by its local interparticle forces and is generally different to that predicted at that location by classical solutions of Boussinesq-like problems. And it is stochastic since it changes with the realization of the experiment. The distribution, value, orientation and ramification of force chains is determined by the features of the system, the boundary conditions and the history of the packing. Although the specific network of force chains resulting after a process cannot be anticipated, the statistics of forces and force chains have attracted considerable attention from the scientific community (e.g.~\cite{Liu95,Jaeger96,Coppersmith96,Radjai96,Mueth98,Radjai99,Majmudar05,Peters05,Radjai15}). Stresses have not been so thoroughly analyzed, while they may be more useful for engineering purposes.\\
	The purpose of this research article is to compare both continuum and discrete approaches and link them through statistics. For example, it  can be intuitively accepted, not yet proven, that the ensemble average~\footnote{An ensemble is an idealization consisting of a large number of virtual copies of the system randomly generated and driven according to the same procedure.} of the stress corresponds to the value given by the solution of the associated boundary value problem (with equivalent values for all intervening parameters).\\
	However not only the mean value of the stresses but also its statistical distribution can be of interest for some applications. In this research,  statistical mechanics principles have been used to anticipate these distributions. Then they have been compared to those measured from simulation. This has been done by numerically generating many packings of a system of discrete elastic particles that are objected to the same macroscopic boundary conditions.\\
	The methodology has been applied to two different problems: the case of a granular half-space under its own weight (Case 1) and a Boussinesq-like problem that is equal to Case 1 but with an additional finite surface load (Case 2).\\
	The work is presented as follows:
	\begin{enumerate}
		\item A theoretical approach to boundary value pro\-blems, stress homogenization techniques and statistical mechanics.
		\item A description of the numerical method and the performed experiments. 
		\item A presentation of results followed with a discussion
		\item An illustrative application. 
		\item A conclusion. 
	\end{enumerate}
	\section{Methods}
	\subsection{Continuum mechanics: the classical solutions}
	\label{sbsc:continuumMechanics}
	\subsubsection{Case1: 2D half-space under its own weight}
	In the absence of any load on the surface, this is just by the action of the gravity, the expected stress field can be determined  from the weight of the overlying material:
	\begin{equation}
	\label{eq:vertStressGravity}
	\sigma_{zz,\text{g}} = \gamma z \text{,}
	\end{equation}
	where $\gamma$ is the unit weight (in kN/m$^3$). $\gamma = \left(1-n \right) \rho_\text{s} \text{g}$, $\rho_\text{s}$ is the density of the material of the particles, $\text{g}$ is the gravitational acceleration and $n$ is the porosity of the packing.\\
	The horizontal stress also increases with depth, but it does at a rate given the at-rest coefficient of lateral earth pressure, $K_0$: $\sigma_{xx,\text{g}}~=~K_0~\gamma z$.\\
	The shear stress is null everywhere $\sigma_{xz,\text{g}}~=~0$, so horizontal and vertical stresses are the principal directions and all the points located at the same depth are thus symmetric.
	\subsubsection{Case 2: 2D Half-space with gravity and a finite surface load}
	For the case of surface loads acting on elastic half-spaces, classical solutions can be revisited. A derivation of the solution to Boussinesq and Flamant problems is found in~\cite{verruijt2007}. In both cases the stress increments caused by the load decrease with the depth but change with the horizontal distance to the applied load. In 2D, the vertical stress in the $x$-$z$ plane caused by a surface load $p$ is given by:
	\begin{equation}
	\label{eq:vertStressFooting}
	\sigma_{zz,\text{p}}=\frac{p}{\pi}\left[ \left(\theta_1 -\theta_2 \right) +\sin{\theta_1}\cos{\theta_1}-\sin{\theta_2} \cos{\theta_2}\right]\text{,}
	\end{equation}
	with $\theta_1=\arctan{(x-X_1)/z}$ and $\theta_2=\arctan{(x-X_2)/z}$ and $X_1$, $X_2$ the left and right limits of the surface load. Close expressions for $\sigma_{xx,\text{p}}$ and $\sigma_{xz,\text{p}}$ can be found in~\cite{verruijt2007}.\\
	However numerical and laboratory experiments with no gravity are difficult. The bearing capacity of the half-space would be too low and the particles on the free surface could fly,   so the hypothesis of small strains would not apply. For that reason, the case in which gravity and surface loads act at the same time has been considered instead in this research. In such circumstaces both solutions (Eqs.~\ref{eq:vertStressGravity}~and~\ref{eq:vertStressFooting}) can be superposed, because the behavior of the material is linear elastic and the equations of equilibrium and compatibility are linear too. Then, $\sigma_{zz}=\sigma_{zz,\text{g}}+\sigma_{zz,\text{p}}$.
	\subsection{Homogenization techniques: from discrete to continuum media}
	\label{sbsc:homogeneitation}
	A heterogeneous continuous body can be partitioned into domains. Every domain $m$ in equilibrium has an inner stress field that matches the solution of the corresponding microscopic elastic problem. In the absence of body forces, the static equilibrium condition is $\sigma_{ij,i} = 0$. The boundary condition at a given point is $\sigma_{ii} n_j = p_j$, with $n_j$, $p_j$ representing the $j$-component of the normal and load vectors, respectively. Considering these two conditions and using the Gauss-Ostrogradsky theorem, the average stress field within the domain $m$ is given by~\cite{Bagi96}:
	\begin{equation}
	\label{eq:averageStress}
	\left\langle \sigma_{ij}^m \right\rangle = \frac{1}{V^m} \int \sigma^m_{ij} dV^m= \frac{1}{V^m} \sum_l x^{mn}_i F^{mn}_j \text{,}
	\end{equation}
	where $V^m$ is the volume of the domain, $F^{mn}_j$ is the $j$-component of the interaction force between domains $m$ and $n$ and $x^{mn}_i$ is the  $i$-component   of the point of application of the force. Equation~\ref{eq:averageStress} is independent of the origin of the coordinate system because particles are in static equilibrium. Therefore, the average stress field within a domain $m$, $\left\langle \sigma_{ij}^m \right\rangle$, can be obtained from the external forces and the positions where they are applied. The tensor product of these vectors is equal to the volumetric average of the stress field multiplied by the volume of the domain $\Sigma_{ij}^m = \left\langle \sigma_{ij}^m \right\rangle V^m$. We will refer to this extensive tensor quantity, $\Sigma_{ij}^m$ , as extensive stress (also known as force-moment), which is expressed in energy units. A very important property of the extensive stress tensor is that it is additive: the extensive stress of a composite body (i.e. its average stress tensor multiplied by its volume) can be obtained either by adding the extensive stress of each separate component or by adding the tensor product of  the external forces on that body  by their position vectors.
	\subsection{Statistical mechanics approaches}
	Statistical mechanics is the branch of physics that deals with systems made of a large number of constituents. Although it was originally developed for thermal systems, it can be applied to granular media. Several approaches~\cite{Edwards89,Edwards03,Edwards05,Henkes07,Henkes09,Tejada2014b} have been proposed since first Edward's model in 1989. That based on the extensive stress~\cite{Henkes07,Henkes09,Tejada2014b} has been followed in this research.\\
	The basic idea of statistical mechanics is~\cite{Balescu1975} that among the solutions of a physical problem (\emph{e.g.} the static arrangement of particles in mechanical equilibrium) there is a class that is compatible to our macroscopic knowledge of the system (\emph{e.g.} it is in equilibrium with some boundary conditions). This class still contains an enormous number of solutions and in the absence of further information there is not any \emph{a priori} reason for favoring one of these more than any other (principle of equal a priori probabilities). This is not a mechanical, but a statistical assumption, because mechanics alone cannot solve the problem uniquely. Furthermore, in equilibrium theory the role of dynamics is trivial: the problem is essentially statistical one.\\
	A half-space made of densely packed particles can be partitioned into domains according to a Voronoi diagram. The volume of each Voronoi domain $V^m$  includes the volume of the elastic particle and an associated void space. The $\mathcal{M}$ domains of the system have inner stress fields whose volumetric average can be obtained from their interaction forces. This length scale is interesting because  this is the scale on which the voids interrupt the continuity of the inner stress field of particles and the volumetric average stress may considerably change from a domain to its adjacent ones.  Each component of the extensive stress of a subsystem $\text{A}$, made of $\mathcal{N}$ domains (with $\mathcal{N} < \mathcal{M}$), is given by $\Sigma^\text{A}_{ij} = \sum_{m=1}^\mathcal{N} \Sigma^m_{ij}$. The volumetric  average stress in the subsystem $\text{A}$ is given by $\left\langle \sigma_{ij}^\text{A} \right\rangle = \Sigma^\text{A}_{ij} / V^\text{A}$, with $V^\text{A} = \sum_{m=1}^\mathcal{N} V^m$.\\
	
\begin{figure}
		\centering
		\includegraphics[width=0.45\textwidth]{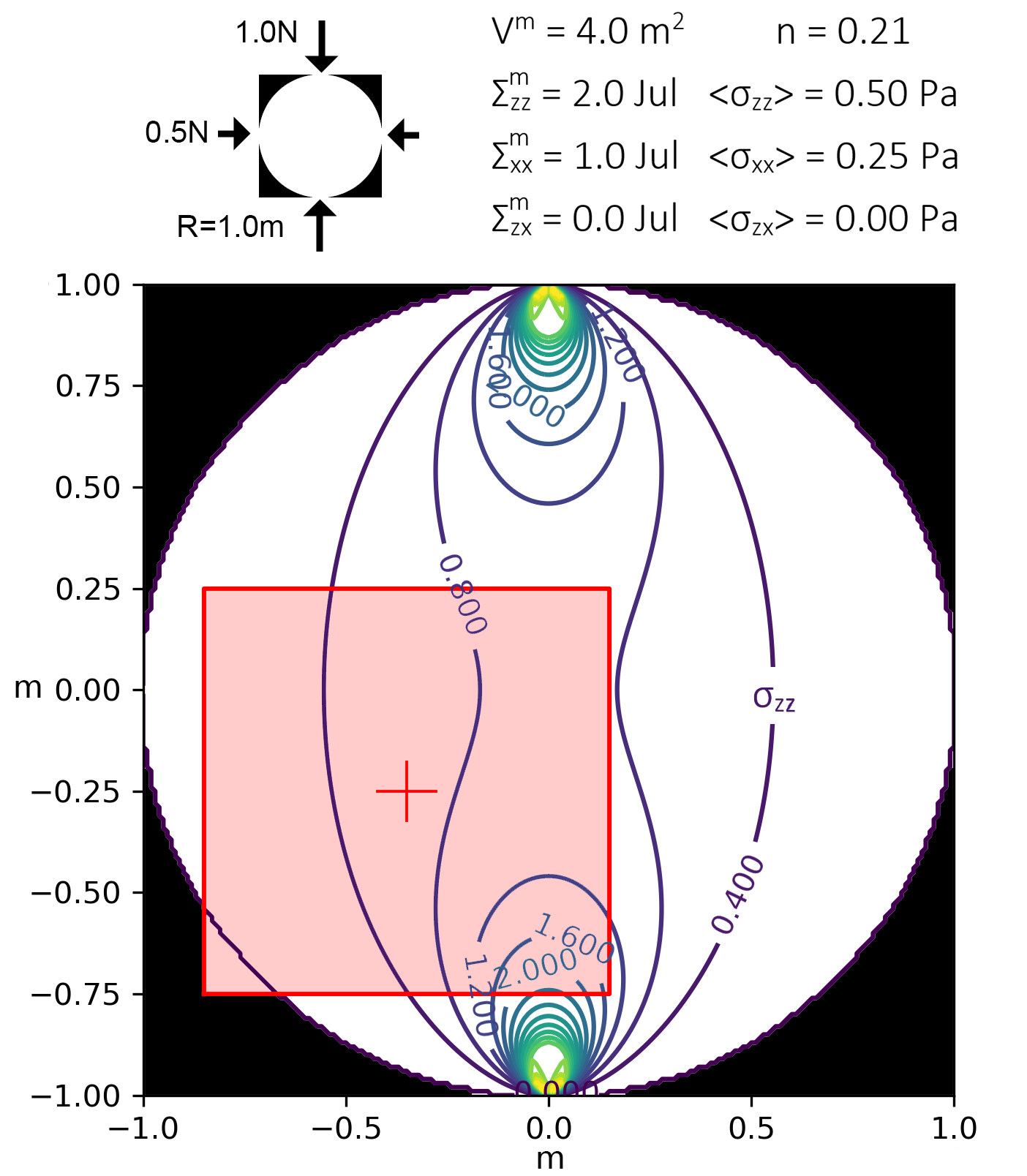}
		\caption{Illustrative example: a particle in static equilibrium and arranged according to a regular simple square lattice. The Voronoi cell is a square of side $2.0$ m. The inner stress field $\sigma_{zz}$, the average stress $\left\langle \sigma_{zz} \right\rangle$, $\left\langle \sigma_{xx} \right\rangle$, $\left\langle \sigma_{xz} \right\rangle$, and the extensive stress $\Sigma^m_{zz}$, $\Sigma^m_{xx}$, $\Sigma^m_{xz}$ caused by two sets of opposite forces of value $1.0$ and $0.5$ N are shown. The area shaded in red is the control region and the black area is the void space.}
		\label{Fig:DiskExample}
	\end{figure}

	The model is based on 4 main hypotheses:
	\begin{enumerate}
		\item[(i)] The volumetric average stress within a control region $c$ of volume $V^c$ extracted from a domain $m$ of volume $V^m$ (with $V^c < V^m$) is equal to the volumetric average of the stress field within the whole domain (see Fig.~\ref{Fig:DiskExample}). This assumption becomes true as $V^c$ approximates $V^m$.  In consequence the extensive stress of the control region is given by $\Sigma_{ii}^c = \Sigma_{ii}^\text{m} V^\text{c} / V^m$. It is worth mentioning that larger particles are associated to larger domains and are more likely selected when measuring the extensive stress at the control point. This is a possible source of bias but the effect is reduced as the particle size distribution concentrates around the mean value (as in this research).
		\item[(ii)] The values of extensive stress of the domains in a packing are independent from each other. Although this is not completely true, the packing is so hyperstatic that the assumption could be acceptable for practical purposes.
		\item[(iii)] In principal directions, the values of normal and shear components are independent from each other. As particles do not break or plastically deform and local arrangements may change the layout and orientation to keep particles in static equilibrium under certain extensive stress conditions, no additional constraint is needed.
		\item[(iv)] The extensive stress tensor of a control region  located at a position $\left( x,z \right)$ may take any allowed value provided that the average value of the statistical distribution is externally controlled by the macroscopic boundary value problem. In other words, if the packing is driven to make forces redistribute, the extensive stress within the control region will surely change but the mean value of the statistical distribution must be equal to that expected from a continuum approach. In principal directions the external control occurs in the next way:
		\begin{itemize}
			\item Normal extensive stresses can take any positive value (negative values are not allowed because tensile interparticle forces do not exist in the interaction model\footnote{There is actually a minimum value for the extensive stress that is given by the own weight of the domain, but this is often negligible when compared to the weight of the overlaying material.}) provided that the mean value matches the corresponding principal stress obtained in a boundary value problem.
			\item Shear extensive stresses can take any positive or negative value provided that the mean value is null and the absolute difference from the mean is limited by interparticle friction and the stress level. 
		\end{itemize}
	\end{enumerate}
	Verifying  these postulates is the object of this research. The statistical samples used on this purpose can be generated by collecting $\mathcal{N}$ control regions that would be objected to identical stress conditions in an equivalent boundary value problem. This is the case of either control regions that are located around the same position in different packings objected to the same constrictions and generated with the same protocol (ensemble sample) or of control regions extracted from the same packing at positions that are supposed to be in similar stress conditions in a macroscopic boundary value problem (packing sample).\\
	
	Following classical statistical mechanics approaches, the data  can be classified into a discrete set of values of the extensive stress  $\left\lbrace \Sigma_{ij,1}, \Sigma_{ij,2}, \cdots, \Sigma_{ij,r}, \cdots \right\rbrace$. The total number of domains with a extensive stress value of $\Sigma_{ij,r}$ is denoted as $\mathcal{N}_{ij,r}$ and the number of permutations of the multiset formed by the extensive stress values $\Sigma_{ij,r}$ is given by the multinomial coefficient:
	\begin{equation}
	\label{eq:multinomial}
	\Omega = {\mathcal{N}\choose \mathcal{N}_{ij,1}, \cdots, \mathcal{N}_{ij,r}, \cdots} = \frac{\mathcal{N}!}{\mathcal{N}_{ij,1}!  \cdots \mathcal{N}_{ij,r}! \cdots} \text{,}
	\end{equation}
	For normal components of the extensive stress tensor ($ij=xx$ or $ij=zz$), the most probable distribution is that maximizing $\Omega$ under the constraints $\sum_r{\mathcal{N}_{ii,r}} = \mathcal{N}$ and $\sum_r{\mathcal{N}_{ii,r} \Sigma_{ii,r}} = \Sigma_{ii} = \mu_{ii}  \mathcal{N} $\footnote{This is actually a simplification since an additional constraint is given by the total volume. This has not considered in this research because the particle size distribution is almost uniform and the variability of local volumes is quite small.}, with $\Sigma_{ii,r} \ge 0$. Then by using natural logarithms, Lagrange multipliers and the Stirling approximation,  the celebrated Maxwell-Boltzmann statistics is found. This statistics states that the probability of finding a domain $m$ with a given vertical extensive stress values is:
	\begin{equation}
	\label{eq:multinomial2}
	\mathcal{P}_{\left( \Sigma^m_{ii} = \Sigma_{ii,r} \right)} = \frac{\mathcal{N}_r}{\mathcal{N}}= \frac{\text{e}^{-\Sigma_{ii,r}/\mu_{ii}}}{\sum_s \text{e}^{-\Sigma_{ii,s}/\mu_{ii}}} \text{.}
	\end{equation}
	If possible values are not discrete but continuously distributed, then an exponential distribution is obtained:
	\begin{equation}
	\label{eq:exponentialBasic}
	f_{\left(\Sigma_{ii} \right)} = \frac{1}{\mu_{ii}} \text{e}^{- \Sigma_{ii} /\mu_{ii} }  \text{.}
	\end{equation}
	In the case of shear stresses, the constraints are  $\Sigma_{xz} \in \left(-\infty, +\infty \right)$ and $E \left(\vert \Sigma_{xz} - 0 \vert \right) = b_{\left( \Phi, \Sigma_{ii} \right)}$, where $b$, the so-called diversity, is a function that grows with interparticle friction and with the stress level.\\
	This statistics only applies on the grain scale. For larger length scales, \textit{i.e.} systems made of several particles, the statistical distribution of the total extensive stress is different, as explained in section~\ref{sc:application}.\\
	
	This theoretical model has been considered in the 2 cases proposed in subsection~\ref{sbsc:continuumMechanics}. In case 1, the statistical distribution of normal extensive stresses in principal directions (which are precisely the horizontal and vertical directions, $x,z$) should be an exponential distribution of mean $\mu_{ii,\text{g}\left( z \right)} = \sigma_{ii} V^\text{c}$, where $V^\text{c}$ is  the volume of the control domain and $\sigma_{ii}$ the stress predicted at that point. For shear stresses, the mean is  $\mu_{xz,\text{g}} =  0$ and values should follow a Laplace distribution of diversity $b$ growing with the depth.\\
	In case 2, the action of the finite surface load breaks the symmetry of the points located at the same depth, so that packing samples are not possible. Nevertherless, at each control point the statistical distributions of normal components are still expected to follow exponential distributions whose means are the value of the corresponding principal stress. In the case of shear stresses the diversity $b$ not only would depend on the depth but also on the horizontal position.
	\subsection{Numerical modeling}
	\subsubsection{The discrete element method}
	The discrete element method~\cite{Cundall79}, implemented in YADE~\cite{yade:reference2} has been used to randomly generate packing compatible to cases 1 and 2. The DEM computes the motion of the solid particles by considering particle-particle interactions. A common frictional-Hookean DEM model was used, so normal interaction forces  grow  linearly with overlaps. The overlap is defined as $\delta_{ij} =  \left( R_i + R_j \right) -\mathbf{r}_{i,j}  =  \left( R_i + R_j \right) - \vert \mathbf{r}_{j} - \mathbf{r}_{i} \vert$,
	where $R_{i,j}$ and $\mathbf{r}_{i,j}$ are the radius and position vector of particles $i$ and $j$, respectively. The normal contact force per unit of lenght acting on particle $i$ due to particle $j$ is:
	\begin{equation}
	\mathbf{F}_{\text{n},ij} = - k_\text{n}  \delta_{ij}  \text{,}
	\end{equation}
	where the contact stiffness is $k_\text{n} = 2 E R_i R_j / \left( R_i + R_j \right)$ (in N.m$^{-1}$) and $E$ is the Young's modulus (in Pa).\\
	Tangential forces are produced in opposition to incremental lateral displacements. These forces are limited by the value of normal forces and friction coefficients. $\mathbf{F}_\text{s}^{ij} = - \min{\left( K_\text{s}  u_{ij}, \tan{\phi} \vert \mathbf{F}_{\text{n},ij} \vert  \right) } \mathbf{u}_{ij} / \vert \mathbf{u}_{ij} \vert$, where $u_{ij}$ is the lateral displacement between the two particles previously in contact ($\delta_{ij} \ge 0$) and $\phi$ is the friction angle and $k_\text{s}$ is an elastic stiffness parameter.
	
	\subsubsection{Numerical experiments} \label{sbsc:numericalExperiments}
	Two sets of numerical experiments were performed (Table~\ref{tab:parameters}). In these sets, the packings were generated by randomly pouring $5000$ particles within a  $1.0$ m wide domain and waiting for an almost complete dissipation of kinetic energy. The sets differ from each other in the interparticle friction angle during the gravity deposition and in the presence or absence of a surface load later on. An additional wider packing (made of $50000$ particles within a $10.0$ m wide domain) was also generated to gather a packing sample in Case 1. The friction in Case 2 is removed during the particle deposition because a dense packing is needed to reach the target load without causing local or punching shear failure modes~\cite{Vesic63}. In all the cases, a quasi uniform particle size distribution was used (\textit{i.e.} all the diameters lying within the interval $D \pm \Delta D$). Gravity acted downwards with $\text{g}=9.81$ m/s$^{2}$. Surface loads were applied in Case 2 by gently and vertically (downwards) moving a rigid body of length $2a$ and centered at $x = 0.0$. These experiments are similar to the punch test carried out by~\cite{Peters05}. Once the total vertical force on this rigid element was equal to $2ap$, the simulation was stoped.\\ 
	Once a packing was in static equilibrium in either Case 1 or Case 2, the statistical distributions of the extensive stress was measured at several control positions. We selected a square control region of volume $V^\text{c}  = 2.5 \cdot 10^{-5}$ m$^2 \simeq D^2/4$. As the position of the control region and the center of the particle used to compute the extensive stress are usually different, some uncertainty is considered. In Case 1 all the particles whose center was located at a height $h_\text{O} \pm \Delta h_\text{O}$ from the bottom in the $10.0$ m wide packing were considered for a packing sample. In both Cases 1 and 2 ensemble samples were generated by collecting of values from control regions around the same point in  many $1.0$ m  wide packings. In Case 1, there was a single point O located right below the center of the box. In Case 2, three control points were considered: point O right below the center of the surface load and points A and B located at the same height that O but that horizontally separate a given distance (leftwards and rightwards) from the center of the load and are not below the rigid body. Points O, A and B were selected because the total stress induced by the surface load there, $\sigma_{zz,\text{p}}$, is quite noticeable, with respective $\sigma_{zz,\text{p}}/\sigma_{zz,\text{g}}$ ratios of $4.59$, $2.15$ and $0.65$. A shear indicator is defined as the ratio of the maximum shear stress to the mean stress $s = \left(\sigma_{1} - \sigma_{3}\right) / \left(\sigma_{1} + \sigma_{3}\right)$. The simulation box was large enough to ensure that the stress field caused by the surface loading is below $0.05p$ at the boundaries.\\
	The average height of the half-space $H$ the porosity of the packing $n$ and the final position of the footing $H_\text{f}$ slightly changed with the realization of the experiment. To measure $H \pm \Delta H$ and $n \pm \Delta n$ a linear regression (Eq.~\ref{eq:vertStressGravity}, with $z=H-h_i$) of the vertical stress computed at different heights $h_i$ was performed (before applying the surface load). The final position of the footing and the actual surface load, with their variation intervals, are directly measured during the experiments. After all these considerations, the uncertainty interval for the expected extensive stress at control points was established. \\
	The properties of the particles used in the simulations are shown in Table~\ref{tab:parameters}.\\
	\begin{table}
		\centering
		\caption{Parameters used in the DEM numerical simulations to generate ensemble samples.}
		\label{tab:parameters}
		\begin{tabular}{lcccl}
			\hline\noalign{\smallskip}
			\multicolumn{2}{c}{ Parameter }
			& \multicolumn{2}{c}{Value} & Units\\
			\noalign{\smallskip}\hline\noalign{\smallskip}
			Number of particles & $N$ & \multicolumn{2}{c}{$5000$} & -\\
			Number of experiments & $\#$ & $1812\,(1)$ & $5324$  &-\\
			Simulation width & $L$ & $1.0\,(10.0)$ & $1.0$  & m\\	
			Mean diameter & $D$ & \multicolumn{2}{c}{$0.01$} & m\\
			Diameter dispersion & $\frac{\Delta D}{D}$ & \multicolumn{2}{c}{$0.05$} & - \\
			Young's modulus & $E$ & \multicolumn{2}{c}{$1.0 \cdot 10^{7}$} & kPa\\	
			Material density & $\rho_\text{s}$ & \multicolumn{2}{c}{$2.6 \cdot 10^{3}$} & kg.m$^{-3}$\\	
			Interparticle friction & \multirow{2}{*}{$\Phi_0$} & \multirow{2}{*}{$\pi/6$} & \multirow{2}{*}{$0$}  & \multirow{2}{*}{rad} \\
			(gravity deposition) & & & & \\
			Interparticle friction & \multirow{2}{*}{$\Phi_1$} & \multirow{2}{*}{-} & \multirow{2}{*}{$\pi/6$}  & \multirow{2}{*}{rad} \\
			(loading) & & & & \\
			Loading width & $2a$ & - & $0.045$ & m\\
			Surface load & $p$ & - & $44.4$ & kPa\\		
			\multirow{2}{*}{Control point O} & $x_\text{O}$ & $0.00$ & $0.00$ & m\\	
			& $h_\text{O}$ & $0.10$ & $0.35$ & m\\
			\multirow{2}{*}{Control point A} & $x_\text{A}$ & - & $0.08$ &m\\	
			& $h_\text{A}$ & - & $0.35$ & m\\
			\multirow{2}{*}{Control point B} & $x_\text{B}$ & - & $0.15$ & m\\	
			& $h_\text{B}$ & - & $0.35$ & m\\
			\noalign{\smallskip}\hline
		\end{tabular}
	\end{table}
	Althouh the displacement of the rigid body during the process was always very small, it is not clear that the behavior of the packing was truly elastic or that it was elastic on both micro and macroscopic levels. In fact, the role of grain level kinematics in the macroscopic behavior  is not fully understood: local fluctuations of potential energy occur~\cite{Sun15} and mesoscale structures form~\cite{Gardinder04}, so that several researches are focusing on this topic -e.g.~\cite{Sibille07,Tordesillas13}-). In fact, in the experiments herein presented, the load was actually applied through a rigid body while the solutions shown in~\ref{sbsc:continuumMechanics} are valid for uniformly distributed loads. However, as the rigid body moves neither the load is uniform nor a flat elastic deformation of the boundary is enforces. The rigid body creates a plastic zone when moving down that  could make it possible to use elastic solutions with uniform loads far away from the plastic region\\
	\section{Results}
	\subsection{Case 1: 2D half-space under its own weight}
\begin{figure}
		\centering
		\includegraphics[width=0.45\textwidth]{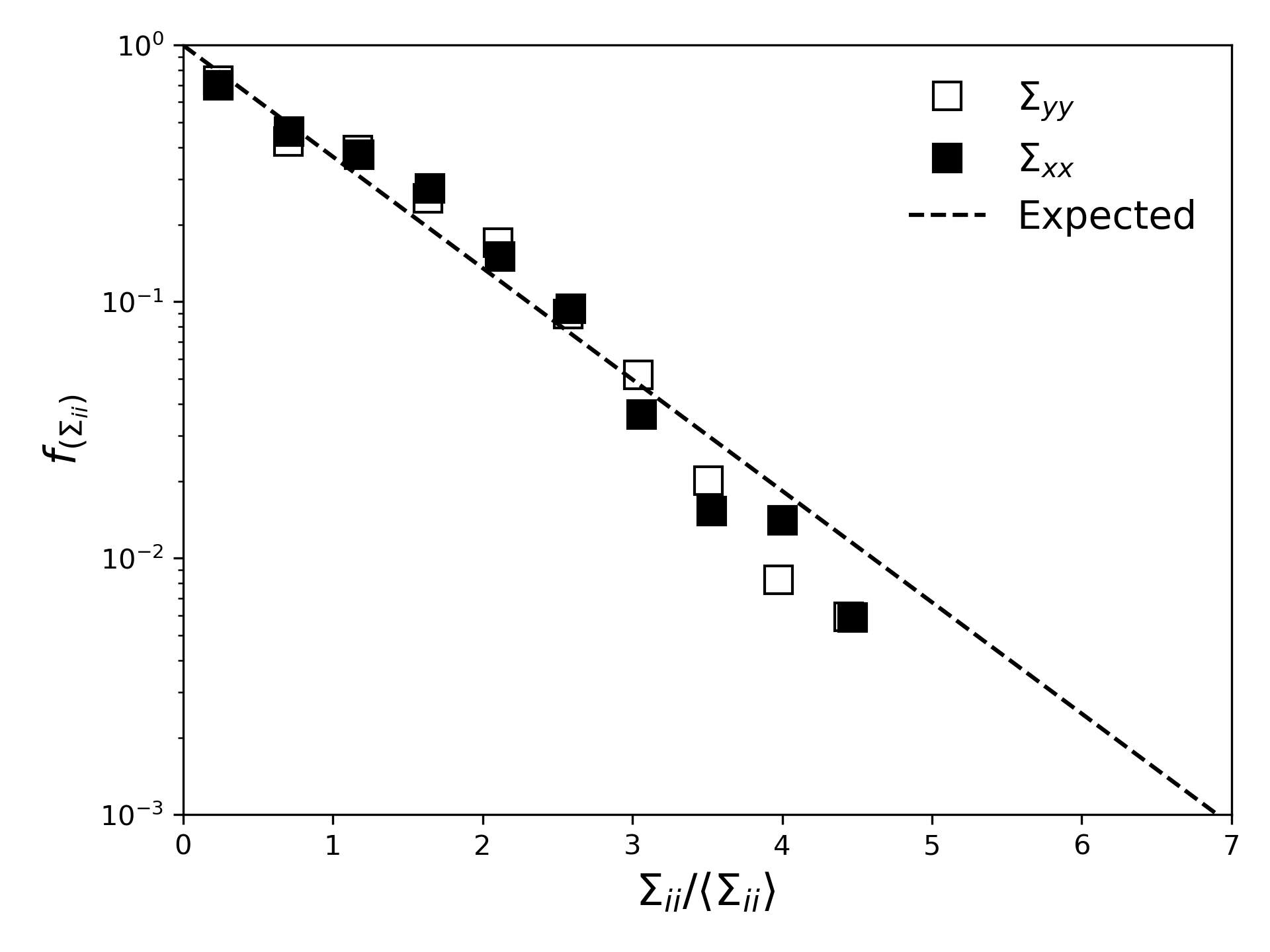}
		\caption{Expected and measured statistical  distribution of normal extensive stresses in Case 1 (data from $1812$ packings).}
		\label{Fig:Case1norm}       
	\end{figure}

\begin{figure}
		\centering
		\includegraphics[width=0.45\textwidth]{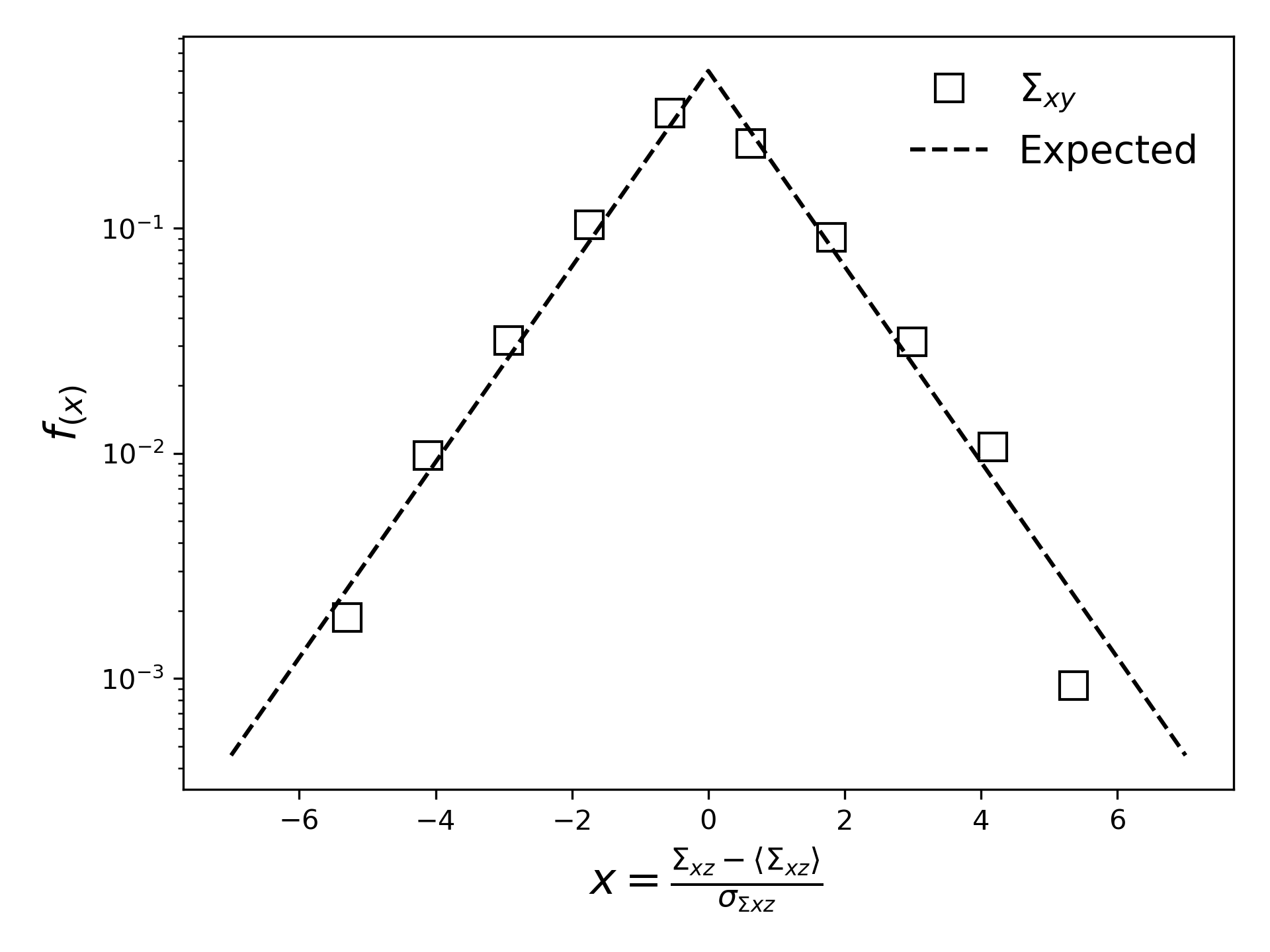}
		\caption{Expected and measured statistical  distribution of shear extensive stress in Case 1 (data from $1812$ packings).}
		\label{Fig:Case1shear}       
	\end{figure}

	The obtained height of the half-space after pouring the particles under gravity action and with interparticle friction angle $\phi=\pi/6$ was $H = 0.49 \pm 0.01$ m. The average porosity, $n = 0.22 \pm 0.01$ and the correlation coefficient of the linear regression was $r^2 = 0.99999$. The at-rest coefficient of lateral earth pressure was  $K_0 = 0.84$, being equivalent to a shear ratio of $s = 0.09$ (almost isotropic compression). The expected vertical stress at the control point  was $7.87 \pm 0.17$ kPa, corresponding to a vertical extensive stress of $(19.68 \pm 0.42) \cdot 10^{-2}$ Jul. The mean value over the ensemble was $19.42 \cdot 10^{-2}$ Jul, perfectly lying within the uncertainty interval. The mean value of the packing sample  was $20.38 \cdot 10^{-2}$ Jul, slightly higher than the upper value of the interval of uncertainty. In the case of horizontal extensive stresses, the expected value was $(16.53 \pm 0.35) \cdot 10^{-2}$ Jul and the sample mean was $16.55  \cdot 10^{-2}$ Jul. \\
	In Figs~\ref{Fig:Case1norm} and~\ref{Fig:Case1shear} the PDFs of extensive stresses of packing and ensemble samples are compared to the expected exponential and Laplace distributions.
	\subsection{Case 2: 2D Half-space with gravity and a finite surface load}
\begin{figure}
		\centering
		\includegraphics[width=0.45\textwidth]{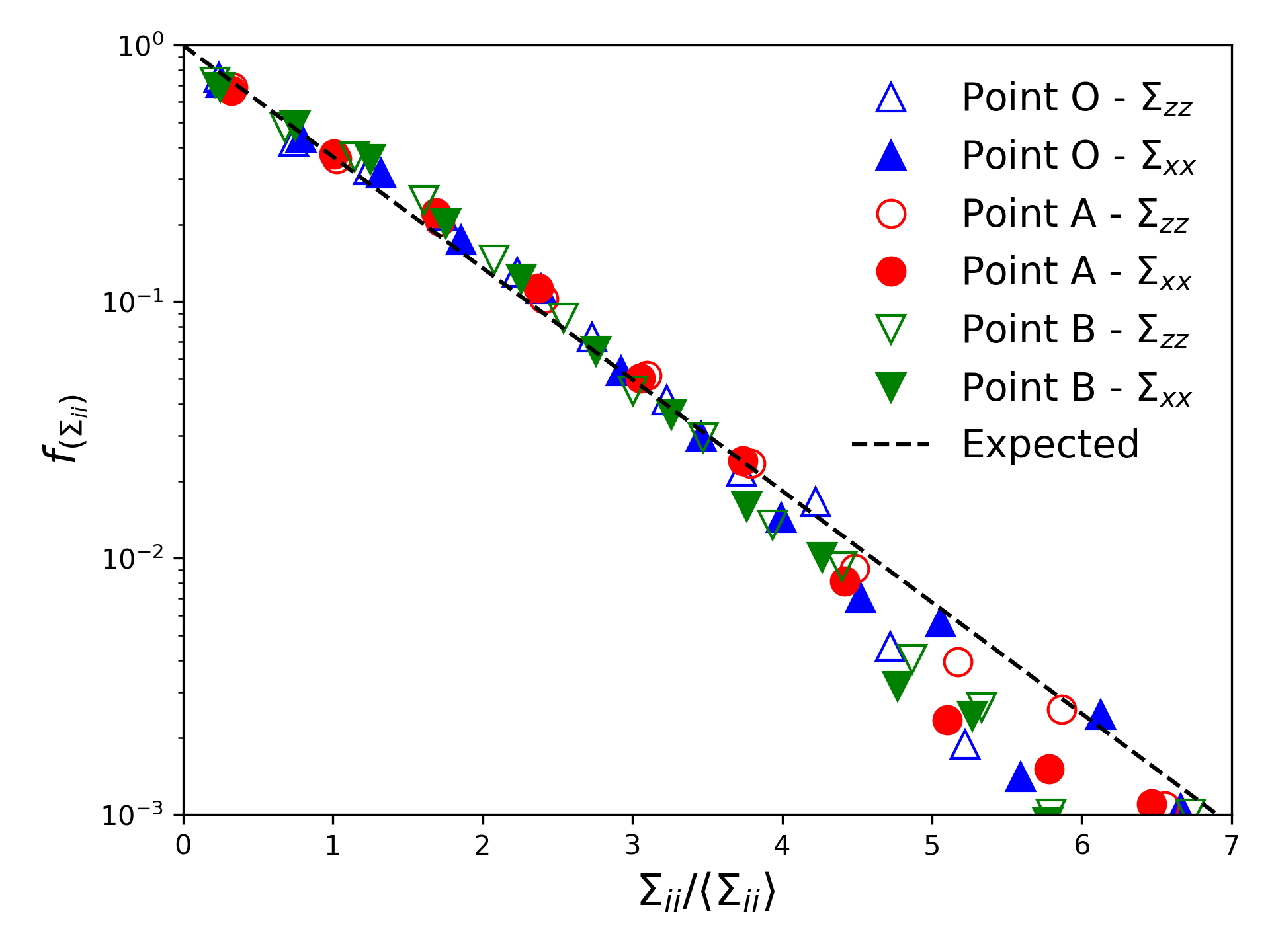}
		\caption{Expected and measured statistical  distribution of normal extensive stresses at 3 different points in Case 2 (data from $5324$ packings).}
		\label{Fig:Case2norm}       
	\end{figure}

\begin{figure}
		\centering
		\includegraphics[width=0.45\textwidth]{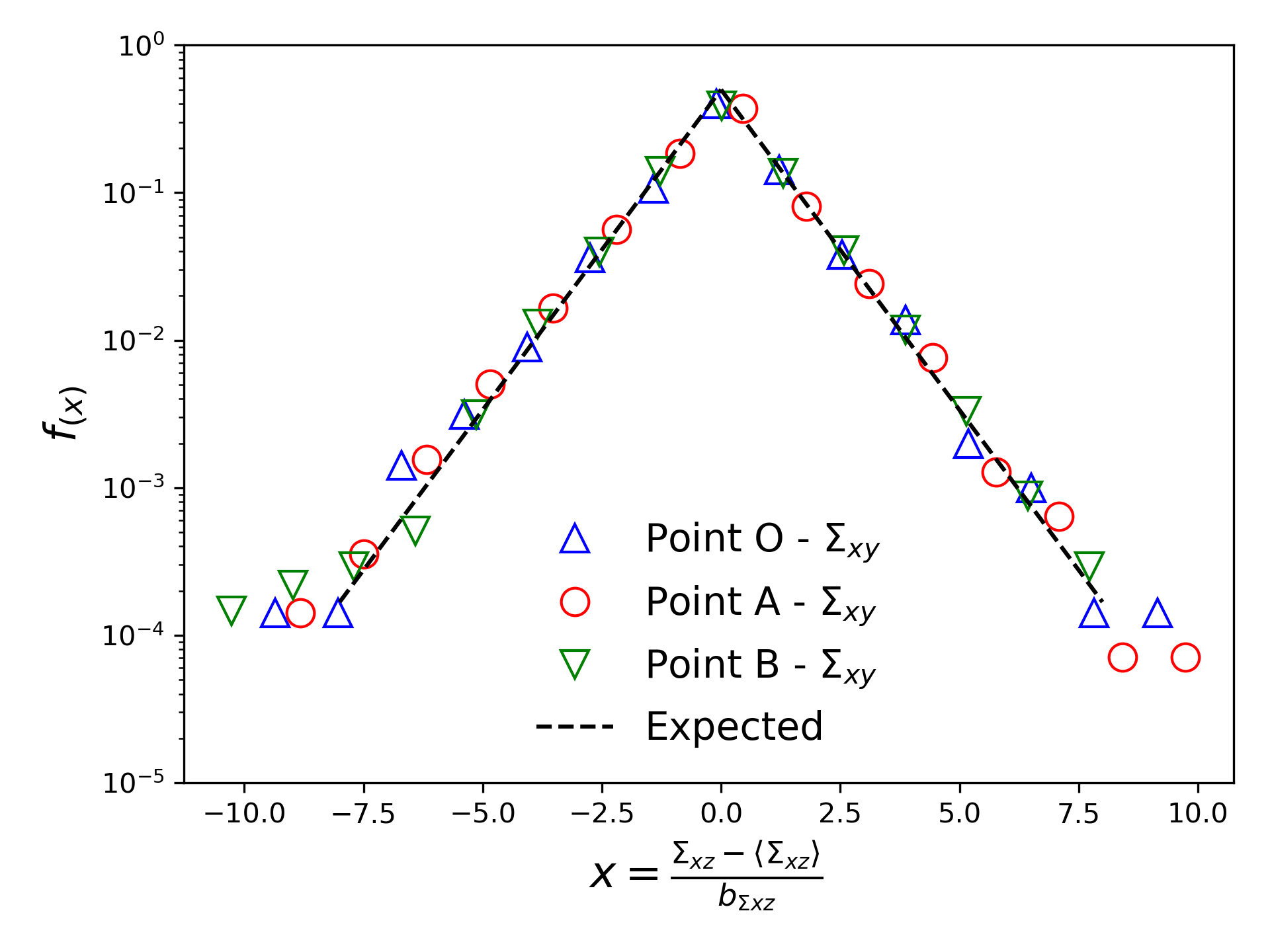}
		\caption{Expected and measured statistical  distribution of shear extensive stress at 3 different points in Case 2 (data from $5324$ packings).}
		\label{Fig:Case2shear}       
	\end{figure}

	In this case particles were frictionless during the gravity deposition stage so they  packed more tightly. The obtained height of the half-space after dropping the particles by gravity was $H = 0.46 \pm 0.02$ m and its average porosity was reduced to $n = 0.15 \pm 0.03$.  After the gravity deposition, $K_0 = 0.95$, with $r^2 = 0.9984$. The load applied by the footing should have increased shear ratios from $s_O = s_A = s_B = 0.023$ to $s_O = 0.693$, $s_A = 0.614$ and $s_B = 0.466$. The principal stresses should have rotated $33.9^{\circ}$ and $51.6^{\circ}$ at points A and B, respectively, and should not have rotated at point O. The expected vertical stresses at points O, A and B were $13.27 \pm 0.40 $ kPa, $7.49 \pm 0.56$ kPa and $3.92 \pm 0.40$ kPa, respectively, and the corresponding expected vertical extensive stresses of the control regions were therefore $(33.17 \pm 1.00) \cdot 10^{-2}$ Jul,  $(18.73 \pm 1.40) \cdot 10^{-2}$ Jul and $(9.81 \pm 1.01) \cdot 10^{-2}$ Jul. The sample means at those points were $32.6 \cdot 10^{-2}$ Jul,  $17.8 \cdot 10^{-2}$ Jul and $8.9 \cdot 10^{-2}$, perfectly lying within the uncertainty interval in all the cases. However the sample mean values of horizontal and shear stresses did not match the values expected from a continuum approach. In fact the mean value of the shear extensive stress was $0$ at the three points, while it should just has been null only at point O. In Figs.~\ref{Fig:Case2norm} and~\ref{Fig:Case2shear} the PDFs of the extensive stresses are plotted. 
	\section{Discussion}
	\label{sc:discussion}
	\paragraph{PDFs of extensive stresses} In both cases 1 (Point O) and 2 (points O, A and B) the statistical distributions of normal stresses follow exponential distributions, as predicted by the proposed model, with no evidence for mismatching in the values below the mean. The fitting with the exponential distribution is indeed better in Case 2 than in Case 1, something that could be related to the higher shear ratios or the larger sample size.\\
	In all the cases and points, shear extensive stresses follow Laplace distributions. The corresponding values of the diversity in each case should have depended only on  the extensive stress level since the interparticle friction angle was the same ($\Phi = \pi / 6$). The extensive stress level can be measured by $P = 0.5 \left( \Sigma_{xx} + \Sigma_{zz} \right)$. A linear regression of the 4 available data results in an intercept very close to $0$ (as expected) and a slope of $0.4093$, which corresponds to a mobilized friction angle of $\Phi^* = 22.2^{\circ}$. This value is smaller than interparticle friction, something reasonable as the interparticle friction imposes the maximum shear forces but $\Phi^*$ has to do with the expected absolute difference from the mean shear stress. Nevertheless, no reason has been found for a linear growth of $b$ with $P$ and the data are not enough to provide better understanding, so this remains as an open question.\\
	\begin{figure}
		\centering
		\includegraphics[width=0.45\textwidth]{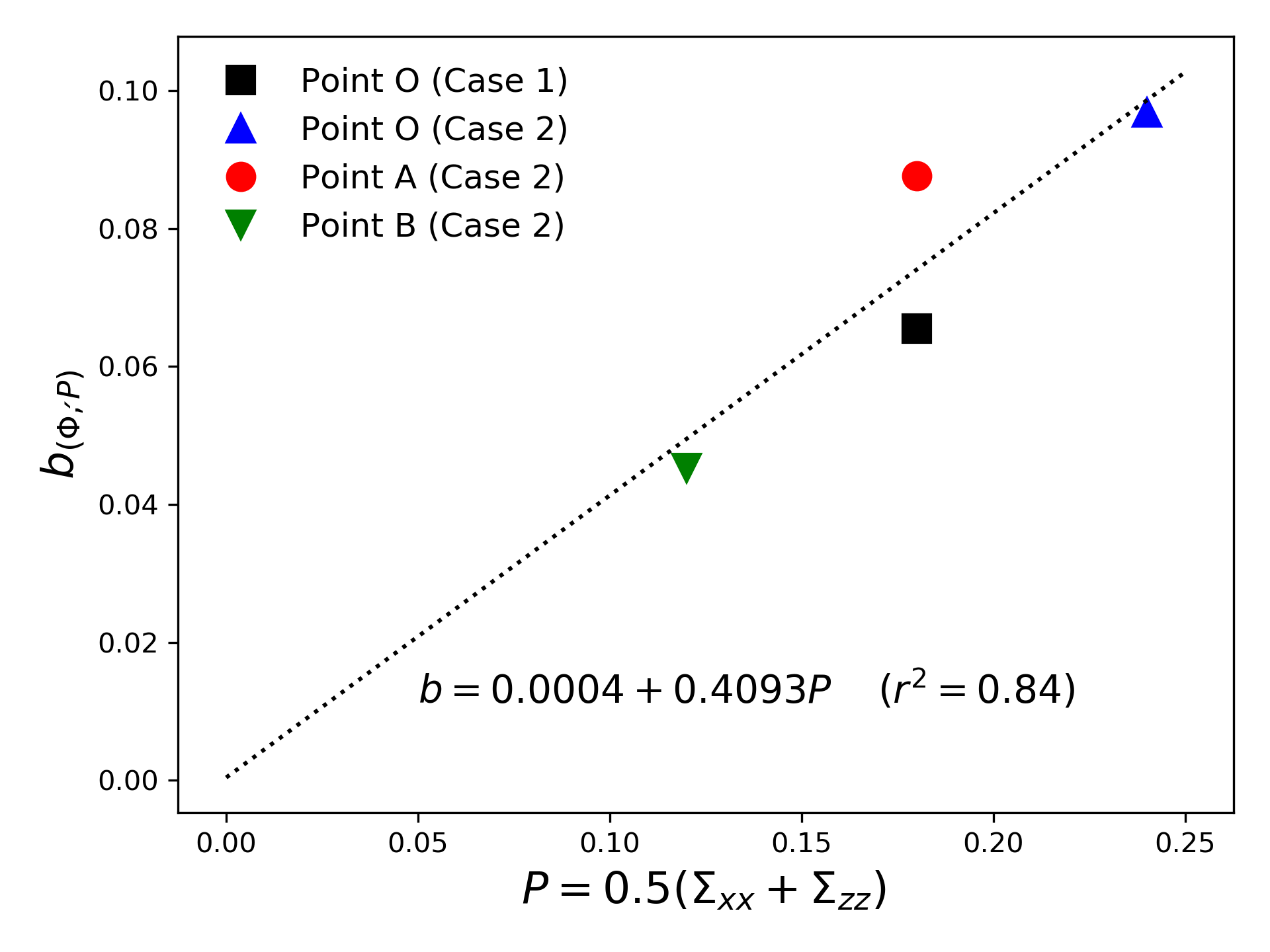}
		\caption{Values of the diversity of Laplace distribution for different values of $P$ in cases 1 and 2 fitted with a linear regression.}
		\label{Fig:bVSp}       
	\end{figure}
	\paragraph{Sample means} The sample means of vertical extensive stress are those expected from a continuum approach (with elastic solutions) in all the cases and control points. However this is not happening for horizontal and shear stresses: in case 1 the values are those predicted by continuum theory but in case 2, the sample means of horizontal and shear extensive stresses are not (except shear extensive stress at point O, which is $0$). This is clearly reflected in the fact that principal stresses keep parallel to horizontal and vertical directions, in opposition to what the elastic problem predicts. This is an issue that deserves more investigation. It could be the consequence of a non elastic process or it could occur because the initial stresses caused by gravity would prevail over those created by the surface load (since shear forces are governed by interparticle friction and particle rotations could occur when the shear resistance is exceeded, balancing the average stress in a different way).\\
	\paragraph{PDFs of forces} As the inhomogeneous transmission of stresses is induced by contact forces, the probability density functions, PDF, of forces (e.g~\cite{Liu95})  and extensive stresses must be related. The celebrated q-model~\cite{Coppersmith96} anticipates a PDF having a peak below the mean value and vanishing with lower forces for a crystalline structure. On the other hand, the analysis of sheared granular materials has provided evidence for a bimodal organization of the force network  into well-defined weak and strong networks~\cite{Radjai96,Radjai99,Radjai15}: the strong network contributes almost exclusively to the shear strength while weak forces act mainly to prop strong force chains. Strong forces follow an exponential decay while weak forces have been found to follow a nearly uniform or power-law decreasing.  However the translation from PDFs of forces into PDFs of stresses is not immediate: the statistics of the contact network (determined by packing history and system features -particle size distribution, friction coefficient, etc.–), the internal constraints (force and moment balance), the strong spatial correlations caused by the complex dynamics that generates mesoscale structures, etc. make this a challenging endeavor. In addition, the PDFs of forces is not always measured in experiments under exactly the same constraints or following the same packing procedure. In fact, the PDF of weak forces seems to be sensitive to the packing state resulting from the deformation history~\cite{Radjai15}. For example, the numerical simulations performed by Radjai and coworkers with different methods in isotropic compaction state and the data obtained by the experiments of~\cite{Mueth98} by means of carbon paper trace coincide everywhere within the available precision in exception to the range of vanishingly small forces. On the other hand, in isotropic packing states the distribution has been found to show a relatively small peak below the mean force (while the probability density of small forces does not fall to zero) but this peak disappears in sheared packings (and the distribution turns to a nearly decreasing power law)~\cite{Radjai96,Radjai99,Radjai15}. In any case, the knowledge on the PDFs of forces could support the exponential decay observed in this research for stresses larger than the average, while it would cast some doubts about smaller values. The history of the packing (with particular influence of the shear level and stress rotation), the features of the system and the stress field imposed by boundary conditions and body forces will finally determine the PDFs of weak and strong forces as well as that of the extensive stress. For engineering purposes, the values higher than the average are often much more useful.\\
	\section{Possible applications}
	\label{sc:application}
	The fact that normal and shear extensive stresses follow exponential and Laplace distributions is a promising finding of this research that could be interesting for geotechnical applications.\\
	As an illustrative example, let be the case of a rigid rectangular framework of width $L$ covered by a granular fill of height $H$ and made of particles of diameter $D$. The stiffness of the framework is equal to that of the surrounding fill. According to continuum approaches the total pressure acting on the top of the framework would be equal to $p = \gamma H $. A direct sampling Monte Carlo simulation was performedto consider the case in which the continuum fill is replaced by a finite number of particles. The vertical extensive stress of those $\mathcal{N} = L/D$ particles interacting with the top of the framework are suposed to follow exponential distributions. For each value of $\mathcal{N}$, $10000$ trials were run and then the variation of the average pressure was evaluated with the coefficient:
	\begin{equation}
	\text{CV} = \frac{p^\text{95} - \bar{p}}{\bar{p}} \text{,}
	\end{equation}
	where $p^\text{95}$ is the 95$^\text{th}$-percentile and $\bar{p}=\gamma H$.
	\begin{figure}
		\centering
		\includegraphics[width=0.45\textwidth]{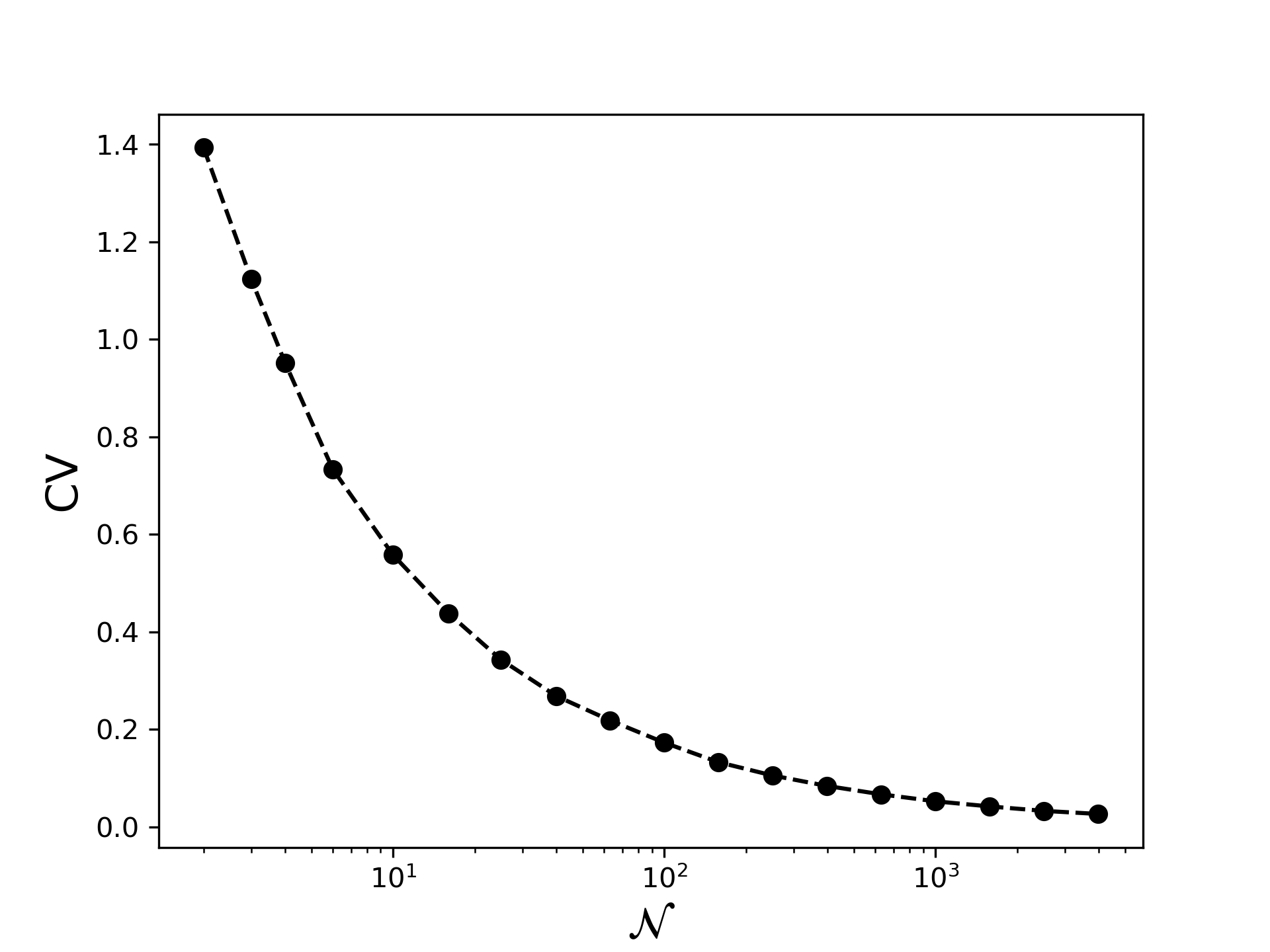}
		\caption{Coefficient of variation of the total pressure acting at the top of a rigid framework under a granular fill.}
		\label{Fig:Example}       
	\end{figure}
	As expected, the variation of resulting extensive stress is more significant as $\mathcal{N}$ decreases. Results are shown in Figure~\ref{Fig:Example}. For example with $\mathcal{N}=L/D=100$, the 95$^\text{th}$-percentile occurs at $1.2p$. It means that with $100$ particles interacting with the roof of the framework, in $5 \%$ of cases the total pressure was $20$\% higher than the mean. A different way to approach this illustrative example, is considering that the average stress is obtained from the contribution of particles whose extensive stress follows exponential distributions. Then the average pressure on the top of the framework follows an Erlang distribution of shape parameter $ \mathcal{N}$ and rate parameter $ \mathcal{N}/(\gamma H)$. The mean value is  $\gamma H$ and the variance $\left( \gamma H \right)^2 / \mathcal{N}$. Then, if the coefficient of variation is defined as the ratio of the standard deviation to the mean value, it decreases with the squared root of $\mathcal{N}$: $C_\text{v} = 1 / \sqrt{\mathcal{N}} $.
	\section{Conclusion}
	A classical statistical mechanics approach has been followed to understand the statistical distribution of the extensive stresses (this is the volumetric average of the stress field of a domain multiplied by the volume of the control region) in two cases: an elastic half-space under its own weight and a elastic half-space under the action of the gravity and of a vertical surface load. Under certain hypotheses, statistical mechanics principles anticipate that the distributions of normal and shear components (in principal directions) are exponential and Laplacian, respectively. Massive DEM simulation has provided evidences of these distributions, with results that are acceptable for practical purposes. In the case of normal extensive stresses, there is no evidence of a different distribution for values lower than the mean, as usually observed in the statistics of interparticle forces. The scaling parameters of these distributions can be predicted by the solution of corresponding boundary value problem in most of the cases but an unexpected missmatching was found in a few of them. This issue remains as an open question.\\
	Although this model has been set up for almost uniform distributions, this is a first step towards the theoretical understanding of the relation between discrete and continuum approaches to geotechnical problems. Anticipating the statistical distributions can be very useful in those situations in which the size of the discrete particles (i.e. the scale of heterogeneities, fragments, etc.) is comparable to the length scale of the problem. The PDFs would, for example, provide the probability of finding stresses that double the values obtained from the corresponding continuum approaches. This research fills a gap between discrete and continuum geotechnical models and opens a way to treat other seminal problems in geotechnics.

\bibliographystyle{plain}
\bibliography{Manuscript}
	
\end{document}